\documentclass[12pt]{article}
\usepackage{amsmath}
 \title{Colored noise influence on the system evolution}
 \author{Dmitrii O.~Kharchenko\thanks{dikh@ssu.sumy.ua}, Sergei Kokhan\\
 }
 \date{ \textit{\begin{footnotesize}Modeling of Complex Systems Dept., Sumy State University, 2,
 Rimskii-Korsakov St, 40007 Sumy, Ukraine\end{footnotesize}} }
\begin{document}
\maketitle
  \begin{abstract}
  We present a picture of  phase transitions of the system with
   colored multiplicative noise. Considering the
  noise amplitude as the power--law dependence of the stochastic  variable
  $x^a$ we show the way to phase transitions disorder--order and
  order--disorder. The governed  equations for
  the order parameter and one-time correlator  are obtained and investigated in details.
  The long--time asymptotes in the disordered
  and ordered domains on  the phase portrait of the system  are defined.

  \textit{Key words:} Phase transitions, Stochastic processes.

\end{abstract}

\section*{I.INTRODUCTION}

Considering an evolution of the stochastic system much more attention is
focused on the statistical moments behaviour. Usually, a system under
consideration is described in terms of order parameter $\eta$, being the first
statistical moment $\langle x\rangle$, where $x$ is a stochastic variable, and
variance $\langle (\delta x)^2\rangle$, which plays role of autocorrelator
($\delta x=x-\langle x\rangle$). Recently it has been shown that a
multiplicative noise can lead a dynamical system to undergo a phase transition
towards an ordered state ($\eta\ne 0$) \cite{VB94,UCNA2}. These results are
obtained for extended systems within Curie--Weiss mean field theory and has
been confirmed through extensive numerical simulations \cite{VB97}. In these
works it was shown that an ordered state exists only inside a window of noise
intensities. Moreover, a spatial distributed system was investigated there and
that kind of phase transition was stipulated by the symmetry breaking. The
existence of the ordered state in the window of the noise intensity was
explained by the collaboration between the multiplicative character of the
noise and the presence of spatial coupling. Here we would like to describe
peculiarities of the noise induced order--disorder phase transition for the
zero--dimensional stochastic system.

If we want to describe the evolution of our system we need to consider temporal
dependencies of the averages according to average motion equation. More
simplest case of ordinary thermodynamic system in this kind approach was
described in \cite{Feldman}, where a kinetic of phase transition is presented.
A system with white multiplicative noise was discussed in \cite{OlKh}, where we
describe the evolution of the system with an arbitrary noise amplitude. Because
the white noise is an idealized model of fluctuations of the system parameters,
here we explore in what a way the colored noise can govern the system
evolution. In this letter, instead of standard approach (small noise spectral
parameter extension), we will use the unified colored noise approximation
developed in \cite{UCNA0,UCNA1,UCNA2} to show the picture of phase transitions
in stochastic systems.

\section*{II.MODEL AND BASIC EQUATIONS}

In the simplest form, the problem of colored noise can be introduced
considering a relevant macrovariable $x(t)$ (density of a given physical
quantity) that satisfies a stochastic differential equation of the form
\begin{equation}\label{1}
\dot{x}=f_0(x)+g_0(x)\lambda(t),
\end{equation}
where $f_0(x)$ represents a deterministic force. A stochastic part of the
evolution is defined by the amplitude $g(x)$ of fluctuations introduced through
the random term $\lambda(t)$, quite often assumed to be gaussianly distributed.
Without loss of generality, the deterministic part of the evolution can be
chosen in polynomial Landau form
\begin{equation}\label{2}
f_0(x)=-\frac{\partial V(x)}{\partial x},\qquad
V(x)=-\frac{\varepsilon}{2}x^2+\frac{1}{4}x^4,
\end{equation}
where $\varepsilon$ is a parameter that acts as the dimensionless temperature,
counted form the critical value.

Considering a whole set of models with a typical behaviour we can use the
power--law function for the noise amplitude
\begin{equation}\label{3}
g_0(x)=x^a,\qquad a\in[0,1].
\end{equation}
This kind of assumption allows us to describe systems with self--similar phase
space and in particular cases we can pass to the ordinary thermodynamic system
($a=0$), directed percolation model ($a=1/2$), population dynamics and forest
fires ($a=1$).

In the simplest case for $\lambda(t)$ we can use definition of
Ornstein--Uhlenbeck process
\begin{equation}\label{4}
\tau\dot{\lambda}=-\lambda+\xi(t),
\end{equation}
where $\tau$ is the correlation time, $\xi(t)$ is the white noise source
($\langle\xi(t)\rangle=0$, $\langle\xi(t)\xi(t')\rangle=\delta(t-t')$).

If we take the time derivative of Eq.(\ref{1}), replace first $\dot\lambda$ in
terms of $\lambda$ and $\xi$ from Eq.(\ref{4}) and then $\lambda$ in terms of
$\dot x$ and $x$ from Eq.(\ref{1}) we can obtain the non--Markovian stochastic
differential equation
\begin{equation}\label{5}
\tau\left(\ddot x-\frac{g_0'}{g_0}\dot x^2\right)+\sigma
\dot x=f_0+g_0\xi(t),
\end{equation}
where
\begin{equation}\label{6}
\sigma=\left[1-\tau\left(
f_0'-f_0\frac{g_0'}{g_0}\right)\right].
\end{equation}
According to the unified colored noise approximation we use
the adiabatic elimination (neglecting $\ddot x$) and
neglect $\dot x^2$.

The problem now lies in obtaining an evolution equation for
the order parameter and autocorrelator. For this purpose we
average reduced equation of motion. It takes the form
\begin{equation}\label{7}
\langle{\sigma(x)\dot{x}}\rangle=\langle{f_0(x)}\rangle.
\end{equation}
The term in right hand side can be represented as a full
derivative of $dy(x)=\sigma dx$, and after averaging,
following \cite{Gardiner}, we get $\langle
dy(x)/dt\rangle=\langle dy(x)\rangle/dt= \langle \sigma
dx\rangle/dt$. Introducing the notation for the
autocorrelator $S=\langle (\delta x)^2\rangle$ we rewrite
right hand side of Eq.(\ref{7}) as
\begin{equation}\label{8}
\langle\sigma(x)\dot{x}\rangle=\dot{\eta}(\epsilon+\kappa(\eta^2+S))+\kappa\eta\dot{S},
\end{equation}
where
\begin{equation}\label{9}
\epsilon=1-\varepsilon\tau(1-a),\qquad \kappa=\tau(3-a).
\end{equation}
The resulting equation for the first statistical moment reads
\begin{equation}\label{10}
[\epsilon+\kappa(\eta^2+S)]\dot{\eta}+\kappa\eta\dot{S}=\eta(\varepsilon-\eta^2)-3\eta
S.
\end{equation}
Because of Eq.(\ref{10}) accounts effect of correlations we
need to construct an equation for the autocorrelator. For
this purpose let us express following differential:
$dy^2=2ydy+(dy)^2$. According to the aforementioned
stochastic process $dy(x(t))$ we find $dy^2=\epsilon^2
dx^2+(2\epsilon\kappa/3)dx^4+(\kappa/3)^2 dx^6$. Making use
of the supposition $x^6\ll 1$ we receive $\langle
 dy^2/dt\rangle=\epsilon^2 d\langle
 x^2\rangle/dt+(2\epsilon\kappa/3)
d\langle x^4\rangle/dt$. Rewriting the Langevin equation in the form of
stochastic differential equation
\begin{equation}\label{11}
{\rm d}y=f_0(x){\rm d}t+g_0(x){\rm d}W
\end{equation}
we immediately produce up to the first order time
derivative (where for Winer process we have $(dW)^2\sim
dt$)
\begin{equation}\label{12}
\begin{split}
 2\epsilon\eta\left[ \epsilon+4\kappa\left(
\frac{1}{3}\eta^2+S \right)\right]\dot{\eta}
+\epsilon[\epsilon+4\kappa(\eta^2+S)]\dot{S}= \\ =2\left[
\epsilon\varepsilon(\eta^2+S)-\left(
\epsilon-\frac{\kappa\varepsilon}{3} \right)
(\eta^4+6\eta^2S+3S^2) \right]+\langle{x^{2a}}\rangle
\end{split}
\end{equation}

The obtained equation combines integer order averages and fractional one,
namely $\langle g_0^2(x)\rangle=\langle x^{2a}\rangle$. As discussed in
\cite{OlKh}, such a kind of fractional average can be expressed in terms of
order parameter and autocorrelator, according to the supposition that the
probability distribution function of the initial process is a homogeneous
function, i.e.
\begin{equation}\label{13}
P(x)\approx Ax^{-2a},\qquad A\equiv
\frac{1}{2}|1-2a|b^{|1-2a|},
\end{equation}
where the cut-off parameter $b\to 0$. One yields the following definition for
the fractional average \cite{OlKh}:
\begin{equation}\label{14}
\langle{x^{nq}}\rangle=\alpha_n(q)\langle{x^n}\rangle^{p_n(q)}
\end{equation}
where
\begin{equation}\label{15}
\begin{split}
p_n(q)&=\frac{1-2a+nq}{1-2a+n},\\
\alpha_n(q)&=A^{\frac{n(1-q)}{1-2a+n}}p_n^{-1}(q)(1-2a+n)^{p_n(q)-1}.
\end{split}
\end{equation}
According to \cite{7} a keypoint of the system with the multiplicative noise
(\ref{3}) is that its behaviour is governed by the magnitude of the exponent
$a$ in Eq.(\ref{3}). At $1/2<a<1$, when the fractal dimension of the phase
space $D=2(1-a)$ is less than 1, the system is always disordered and its
evolution is represented by the the autocorrelator $S(t)$. It provides $q>1$ in
Eq.(\ref{14}), hence the fractional order average is represented by $S(t)$. In
the case $a<1/2$, where $D>1$, according to the Landau theory the system can
test the phase transition and in the fractional order average we account more
essential contribution given by the order parameter $\eta$ ($q<1$ in
Eq.(\ref{14})). Therefore, we can rewrite Eq.(\ref{14}) as
\begin{eqnarray}\label{16}
\langle{x^{2a}}\rangle=\left\{ \begin{array}{ll}
\alpha_1\eta^{p_1}:\quad & 0<a<\frac{1}{2}
\\ \alpha_2S^{p_2}:\quad & \frac{1}{2}<a<1
\end{array} \right.
\end{eqnarray}
where
\begin{equation}\label{17}
\begin{split}
\alpha_1=A^{(1-2a)p_1}p_1^{-p_1},\quad p_1=\frac{1}{2(1-a)},\\
\alpha_2=A^{2(1-a)p_2}p_2^{-p_2},\quad p_2=\frac{1}{(3-2a)}.
\end{split}
\end{equation}

\section*{III.EVOLUTION OF DISORDERED SYSTEM}

Let us analyze the evolution of the disordered system, at first. Considering
the case $a>1/2$ we get $\eta(t)=0$, hence the system evolution is defined by
solutions of the following equation
\begin{equation}\label{18}
\dot{S}\left(\frac{\epsilon}{2}+2\kappa S\right)=S\left(\varepsilon
-S\left(3-\frac{\varepsilon\kappa }{\epsilon}\right)\right)+\alpha_2S^{p_2}.
\end{equation}
The form of time dependencies for the autocorellator is
shown in Fig.1.
It is seen that $S(t)$ monotonically attains the stationary magnitude
determined by the equation
\begin{equation}\label{19}
\varepsilon
-\left(3-\frac{\varepsilon\kappa}{\epsilon}\right)S_0+\alpha_2S_0^{p_2-1}=0
\end{equation}
at condition $S_0\ne-\epsilon/4\kappa$. In Fig.2 we plot
steady states at different values of noise correlation
times and at different values of the control parameter
$\varepsilon$.
 According to Eq.(\ref{19}) with
$\varepsilon$ or $\tau$ increase the stationary value $S_0$
rises from the minimal magnitude. Let us focus on the limit
$S\ll 1$. If we put $S^{p_2}\gg S\gg S^2$ Eq.(\ref{18})
gives the power--law time dependence
\begin{equation}\label{20}
S_{t\rightarrow 0}=Bt^\frac{1}{1-p_2},\quad B=\left(
\frac{2(1-p_2)\alpha_2}{\epsilon}\right)^{\frac{1}{1-p_2}}
\end{equation}
In the opposite case $S_0-S\ll S_0$ one has exponential dependence
$S-S_0\propto e^{\lambda t}$, where $\lambda=
(\varepsilon-S_0(3-\varepsilon\kappa/\epsilon)+\alpha_2p_2)/(\epsilon/2+2\kappa
S_0)$.

\section*{IV.EVOLUTION OF ORDERING SYSTEM}
We now in a position to discuss the case $a<1/2$ of ordering system with the
fractal dimension in $x-t$ space $D>1$. It provides that the system dynamics is
governed by equations for the order parameter and autocorrelator
\begin{eqnarray}\label{21}
&&\gamma(\eta, S)\dot{\eta}=\nonumber \\
 &=&\eta[\epsilon-\eta^2-3S][\epsilon+4\kappa(\eta^2+S)]\\
 &-&2\kappa\eta \left[\varepsilon(\eta^2+S)-\left(1-\frac{\kappa\varepsilon}{3\epsilon}
\right)(\eta^4+6\eta^2S+3S^2)\right]\nonumber \\
 &-&\kappa\epsilon \alpha_1\eta^{p_1+1},\nonumber
\end{eqnarray}
\begin{eqnarray}\label{22}
\beta(\eta, S)\dot{S}&=& \nonumber \\ &=&\left[\varepsilon(\eta^2+S)-\left(
 1-\frac{\kappa\varepsilon}{3\epsilon}\right)
 (\eta^4+6\eta^2S+3S^2)\right] \nonumber \\
 &\times& [\epsilon+\kappa(\eta^2+S)] \nonumber \\
 &-& \eta^2\left[\epsilon+4\kappa\left(\frac{\eta^2}{3}+S
\right)\right][\varepsilon-\eta^2-3S] \nonumber \\
&+&[\epsilon+\kappa(\eta^2+S)]\alpha_1\eta^{p_1}, \nonumber
\end{eqnarray}
where
\begin{eqnarray}
\gamma(\eta, S )&=&[\epsilon+\kappa(\eta^2+S)][\epsilon+4\kappa(\eta^2+S)]\\
&&-2\eta^2\kappa\left[\epsilon+4\kappa\left(\frac{\eta^2}{3}+S
\right)\right],\nonumber
\end{eqnarray}
\begin{eqnarray}
\beta(\eta, S )&=& \left[\frac{\epsilon}{2}+2\kappa(\eta^2+S)\right]
[\epsilon+\kappa(\eta^2+S)]\\
&&-\kappa\eta^2\left[\epsilon+4\kappa\left(\frac{\eta^2}{3}+S
\right)\right].\nonumber
\end{eqnarray}

The obtained closed-loop system of differential equations can be analyzed with
a help of the phase plane method. From the corresponding phase portrait shown
in Fig.3a it is seen that at small values of the control parameter
$\varepsilon$ there is only one attractive point $C_0$ with coordinates
$\eta_0=0$,
$S_0=(\varepsilon/3)[(1-\varepsilon\tau(1-a))/(1-2\varepsilon\tau(1-2a/3))]$.
Increasing of the control parameter provides the appearance
of saddle and attractive points with coordinates given by
solutions of the stationary equations
\begin{eqnarray}
 [\epsilon&-&\eta_0^2-3S_0][\epsilon+4\kappa(\eta_0^2+S_0)]=\\
 &=&2\kappa \left[\varepsilon(\eta_0^2+S_0)-\left(1-\frac{\kappa\varepsilon}{3\epsilon}
\right)(\eta_0^4+6\eta_0^2S+3S_0^2)\right]\nonumber \\
 &+&\kappa\epsilon \alpha_1\eta_0^{p_1},\nonumber
\end{eqnarray}
\begin{eqnarray}
&&\left[\varepsilon(\eta_0^2 + S_0)-\left(
 1-\frac{\kappa\varepsilon}{3\epsilon}\right)
 (\eta_0^4+6\eta_0^2S_0+3S_0^2)\right]\times \nonumber \\
 &\times& [\epsilon+\kappa(\eta_0^2+S_0)]=  \\
 &=&
 \eta_0^2\left[\epsilon+4\kappa\left(\frac{\eta_0^2}{3}+S_0
\right)\right][\varepsilon-\eta_0^2-3S_0] \nonumber \\
&-&[\epsilon+\kappa(\eta_0^2+S_0)]\alpha_1\eta_0^{p_1}. \nonumber
\end{eqnarray}
The dependence of the bifurcation value of the control
parameter $\varepsilon_0$ from the noise correlation time
$\tau$ is shown in Fig.4.
 It illustrates the appearance of the ordered phase domain
($\eta_0\ne 0$) at small values of the noise correlation time and small values
of $\varepsilon_0$. So the ordered phase is realized in the window of the
intensities $[\varepsilon_0^1, \varepsilon_0^2]$. Increasing of $\tau$ provides
the decreasing of the domain of ordered phase. Dependencies of the steady
states are shown in Fig.5,6. Here thin lines display the saddle point $S$ and
thick lines correspond to the attractive point $C$ in Fig.3b. Some distinctive
feature can be seen from Fig.5,6: the system undergoes the phase transition of
the first order despite the bare $x^4$-potential corresponds to the continuous
one. Such a kind of transition can be observed at disorder--order and
order--disorder phase transitions with increasing and decreasing of the control
parameter $\varepsilon$.

Let us discuss now time dependencies corresponding to the phase trajectories
from Fig.3. Time dependencies are shown in Fig.7.

Thus, according to the phase portraits shown in Fig.5, at
$\varepsilon\ne[\varepsilon_0^1,\varepsilon_0^2]$ the order parameter falls
down monotonically to the point $C_0$, whereas the autocorrelator can vary
nonmonotonically. Inside the domain $[\varepsilon_0^1,\varepsilon_0^2]$, where
bifurcation occurs, we can see two domains on the phase plane. These domains
correspond to the small and large values of the order parameter. At small
initial values of the order parameter we receive above mentioned behaviour. At
intermediate and large magnitudes the order parameter attains the attractive
point $C$. In the vicinity of the saddle point $S$ we get a critical
slowing-down and metastable phase can exists some short time. Such a kind of
behaviour can be found near the separatrix $C_0SC$.

Let us analyze the time dependencies of main averages
analytically. Firstly, we investigate the system evolution
in the disordered state (i.e. vicinity of the point $C_0$)
for large time. According to the \cite{OlKh} instead of the
ordinary Lapunov method we use generalized Tsallis exponent
\cite{Tsallis1}:
\begin{equation}
e^{qt}\to \exp_{ q}(t)\equiv [1 + (1 - q) t]^{1/1-q}
\label{a0}
\end{equation}
with generalized Lyapunov index $q$. This construction is more convenient to
analyze the power--law dependencies that we have in evolution equations.
According to the derivation rule
\begin{equation}
\frac{\partial}{\partial t} \exp_{ q}(t) =\left(\exp_{
q}(t)\right)^q \equiv \exp_{q}^q (t) \label{b0}
\end{equation}
and asymptotic behaviour
\begin{equation} \lim_{t\to 0}\exp_{ q}(t)\to 1+t,\qquad
\lim_{t\to\infty}\exp_{ q}(t)\to
\left[(1-q)t\right]^{1/1-q} \label{c0}
\end{equation}
let us assume solutions of Eqs.(\ref{21},\ref{22}) in the
form
\begin{equation}
\eta(t)=m\exp_{\mu}(t), \quad S(t)=S_0+n\exp_{\nu}(t). \label{a1}
\end{equation}
Inserting Eq.(\ref{a1}) into Eq.(\ref{21}) we receive up to the first order of
$m,n\ll 1$:
\begin{equation}
 (\epsilon+\kappa S_0)(\epsilon+4\kappa S_0)=
 -\kappa\epsilon\alpha_1m^{p_1}\exp_\mu^{p_1+1-\mu}(t),
\end{equation}
where we account the singular contribution only. In the long--time limit, the
 function  $\exp_\mu^{p_1+1-\mu}(t)$ can be taken equal to 1. It provides definition
of the Lyapunov multiplier and exponent
\begin{eqnarray}
  m&=&\left|(\epsilon+\kappa S_0)(\epsilon+4\kappa S_0)/\kappa\epsilon\alpha_1\right|^{1/p_1}, \\
  \mu&=&1+\frac{1}{2(1-a)}.\nonumber
\end{eqnarray}

Considering behaviour of the autocorrelator $S(t)$ up to the first order of
amplitudes $m$ and $n$ we obtain
\begin{equation}
(\epsilon/2+2\kappa S_0)=\exp_\nu^{1-\nu}\left(
 \varepsilon+\alpha_1 n^{-1}m^{p_1}\exp_{\mu}^{p_1}(t)\exp_{\nu}^{-1}(t)\right).
\end{equation}
In the short-time limit we assume $\exp_\nu^{1-\nu}\to 1$. The long-time
asymptote yields $\exp_{\mu}^{p_1}(t)\exp_{\nu}^{-1}(t)=const\equiv p_1^{-1}$
and provides
\begin{equation}
\nu=2,\qquad n=\frac{1}{\kappa\epsilon p_1}\frac{(\epsilon+\kappa
S_0)(\epsilon+4\kappa S_0)}{\epsilon/2-\varepsilon+2\kappa S_0}.
\end{equation}
Thus, according to the obtained time dependencies, we see that the order
parameter behaves itself in a power--law form, i.e. $\eta(t)\propto
t^{-2(1-a)}$, and the autocorrelator $S(t)$ follows to the hyperbolical
dependence $S(t)\propto t^{-1}$.

If we pass through the critical value $\varepsilon_0^1$ then the system can be
ordered and here we have to account an initial magnitude of the order
parameter. Taking the initial magnitude of $\eta(0)$ larger then a critical
value $\eta_c$ (shown in Fig.8) we make the system to be ordered.
Let us examine the system evolution to the ordered state (vicinity of the point
$C$). Here we have to note, that we may not use the solution like generalized
exponent (\ref{a0}). The latter is applicable at non--linearity effects which
are sufficient to fix the amplitudes $m$ and $n$. In the case under
consideration the linear conditions are satisfied and because of power--law
construction of the evolution equations we ought to use the Mellin
transformations
\begin{equation}\begin{split}
\eta(t)=\eta_0+\int m_q t^q{\rm d}q,\\ \label{q1} S(t)=S_0+\int n_q t^q {\rm
d}q. \end{split}
\end{equation}
Here the evolution equations can be transformed to the system of linear
algebraical equations
\begin{equation}\label{M1}
\begin{split}
 A_{11}m_q+A_{12}n_q=0,\\
 A_{21}m_q+A_{22}n_q=0,\\
\end{split}
\end{equation}
where multipliers $A_{ij}$ are functions of the system parameters and
coordinates of the point $C$. The diagonal elements of the matrix $\mathbf{A}$
incorporate terms $q/t$, so we can rewrite $A_{ii}=A^{(0)}_{ii}+q/t$. The
system (\ref{M1}) has solution when $\det|\mathbf{A}|=0$. So if we use the
following notation $c=-q/t$ we get
\begin{equation}
\begin{split}
 \eta(t)=\eta_0+m\exp(-ct\ln(t)),\\
 S(t)=S_0+n\exp(-ct\ln(t)),
\end{split}
\end{equation}
here amplitudes $m,~n$ correspond to the index $q=-ct$; $c$ is the real number
whose magnitude can be expressed
 from the condition $\det|\mathbf{A}|=0$:
\begin{equation}
c=\frac{1}{2}(A_{11}^{(0)}+A_{22}^{(0)})\left( 1\pm \sqrt{1-
 \frac{4(A_{11}^{(0)}A_{22}^{(0)}-A_{12}A_{21})}
{(A_{11}^{(0)}+A_{22}^{(0)})^2}} \right).
\end{equation}

\section*{V.SUMMARY}

We have studied dynamics of noise induced order--disorder phase transition.
Colored multiplicative noise contributes to realizing of two phase transitions
with the control parameter increasing. We can conclude that this kind
transitions are a consequence of collaboration of the nonlinearity of the
system and multiplicative character of the colored noise (in the white noise
limit there is only one phase transition along the axis $\varepsilon$
\cite{OlKh}). We have shown that color of the noise does not change time
asymptotes in the vicinity of the attractive points, corresponding to the
disordered and ordered domains, comparing to the white noise limit. It changes
only amplitudes of the time dependencies. We believe that these results may be
found in physical and biological systems possessing a self-affine phase space.

\newpage
\centerline{\textbf{\Large{FIGURE CAPTIONS}}}
\begin{description}

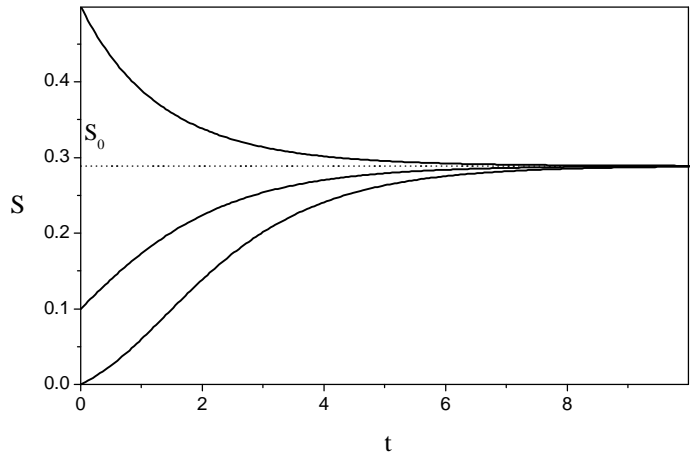
\item[Fig.1] Time dependence of the autocorrelator $S$ at $\varepsilon=0.6$, $\tau=0.5$, $a=0.8$

\item[Fig.2] Stationary states of the system at $a=0.8$: (a) $S_0$ vs. control parameter at different values of $\tau$;
 (b) $S_0$ vs. noise correlation time at different values of $\varepsilon$.

\item[Fig.3] Phase portraits at $a=0.2$: (a)  $\varepsilon=0.1$, $\tau=0.01$; (b) at $\varepsilon=0.5$, $\tau=0.2$.

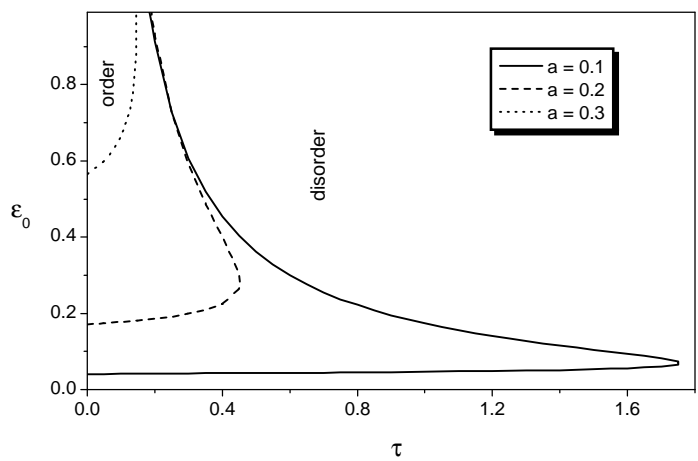
\item[Fig.4] Phase diagram.

\item[Fig.5] Stationary states of the system at $a=0.2$: (a)
 order parameter  vs. control parameter $\varepsilon$;  (b) order parameter  vs. noise correlation time.

\item[Fig.6] Stationary states of the system at $a=0.2$: (a)
 autocorrelator  vs. control parameter $\varepsilon$ (b)  autocorrelator  vs. noise correlation time.

\item[Fig.7]  Time dependencies correspond to the different trajectories on the phase portrait in Fig.3:
 (a) $\eta$ vs. $\ln(t)$ at $a=0.2$, $\varepsilon=0.5$, $\tau=0.2$; (b) $S$ vs. $\ln(t)$ at $a=0.2$, $\varepsilon=0.5$, $\tau=0.2$.

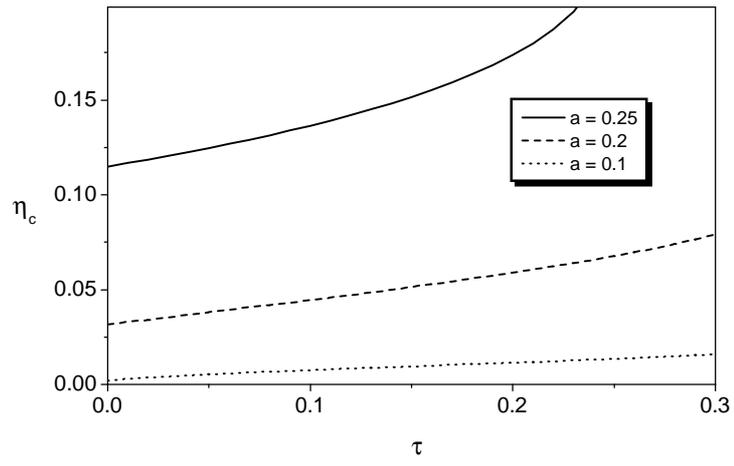
\item[Fig.8] Critical value of the order parameter $\eta_c$ vs. noise correlation time $\tau$ at $\varepsilon=0.5$ and different values of the noise exponent $a$.
\end{description}

\thebibliography{00}
\bibitem{VB94} C.Van der Broeck, J.M.R.Parrondo, R.Toral, Phys.Rev.Lett,
\textbf{73}, 3395, (1994).
\bibitem{UCNA2}S.E.Mangioni, R.R.Deza, R.Toral, H.S.Wio, Phys.Rev.E,
\textbf{61}, 223, (2000).
\bibitem{VB97} C.Van der Broeck, J.M.R.Parrondo, R.Toral, R.Kawai, Phys.Rev.E,
\textbf{55}, 4084, (1997).
\bibitem{Feldman} L.I.Stefanovich, E.P.Feldman,  JETP, \textbf{113},
228, (1998).
\bibitem{OlKh}A.I.Olemskoi, D.O.Kharchenko, Physica A, \textbf{293}, 178, (2001).
\bibitem{UCNA0} P.Jung, P.H\"anggi, Phys.Rev.A \textbf{35}, 4467, (1987).
\bibitem{UCNA1}F.Castro, H.S.Wio, G.Abramson, Phys.Rev.E
\textbf{52}, 159, (1995).
\bibitem{Gardiner} C.W.Gardiner, \textit{Handbook of Stochastic
Methods}(Springer-Verlag, Berlin, 1983).
\bibitem{7}  A.I.~Olemskoi, D.O.~Kharchenko, Met.Phys.Adv.Tech., {\bf
16}, 841 (1996).
\bibitem{Tsallis1} C.~Tsallis,
in {\it Nonextensive Statistical Mechanics and its
Applications}, {\it Lecture Notes in Physics}, eds. S. Abe
and Y. Okamoto (Springer-Verlag, Berlin, 2000).
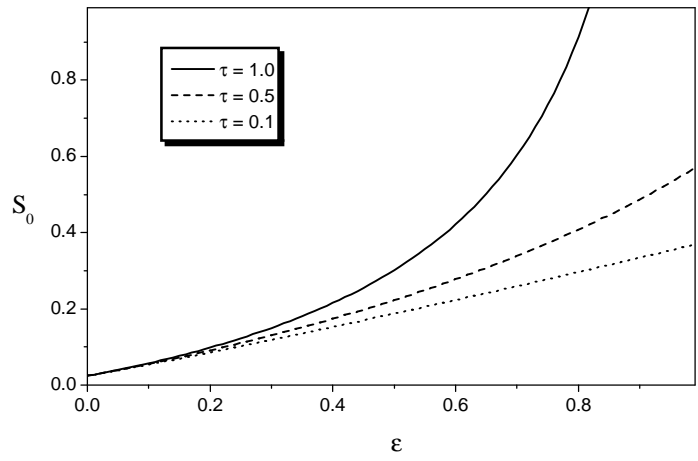
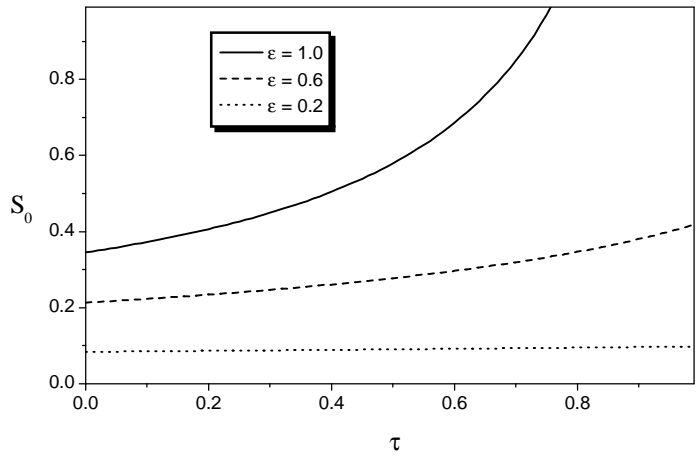
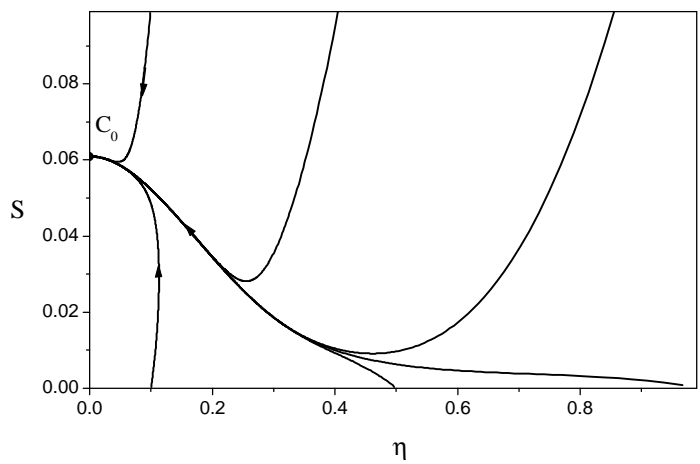
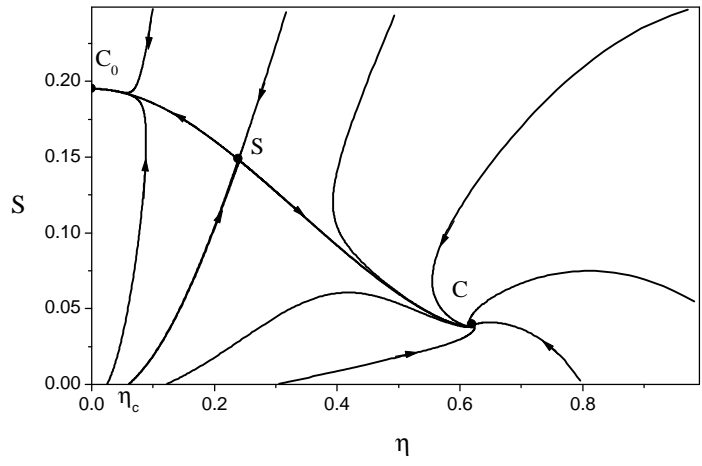
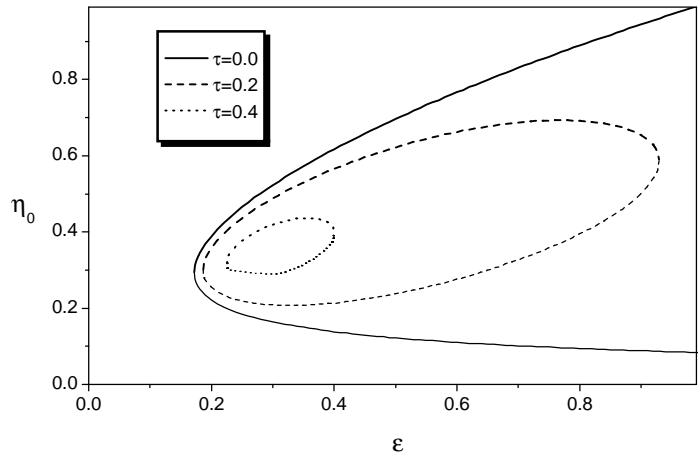
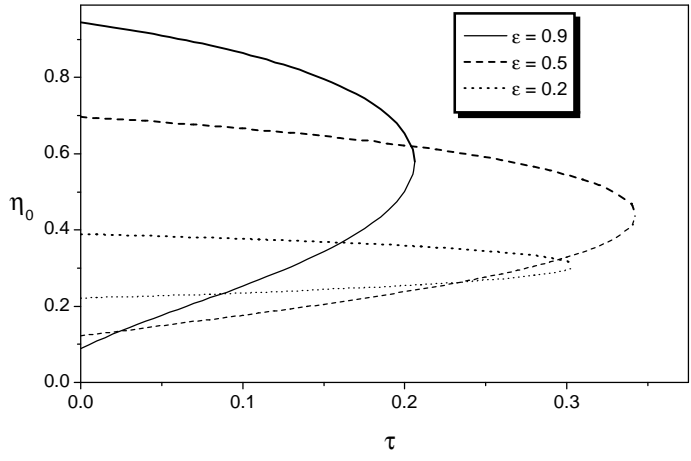
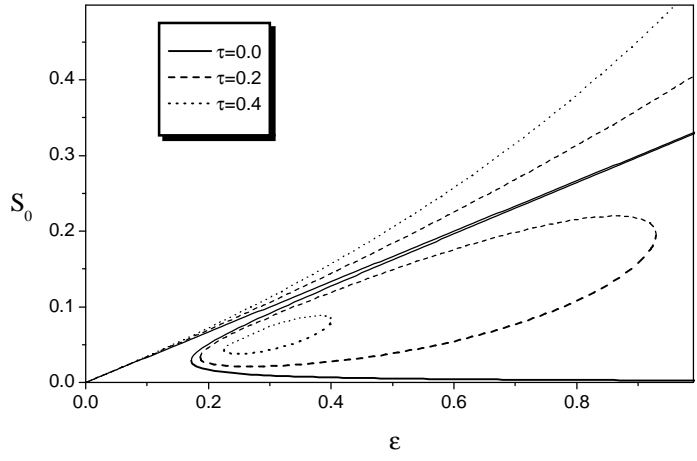
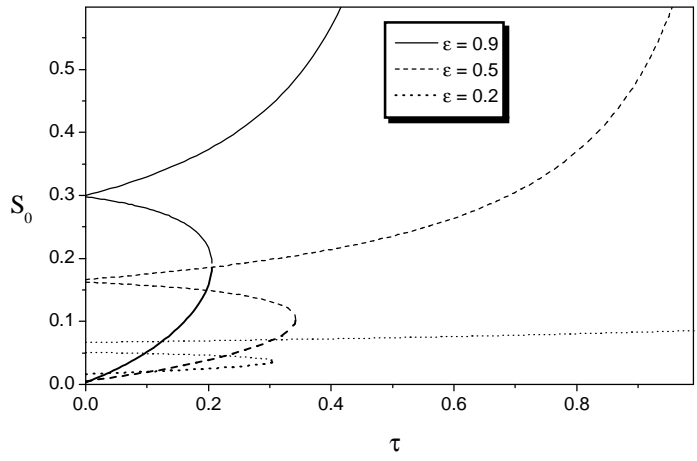
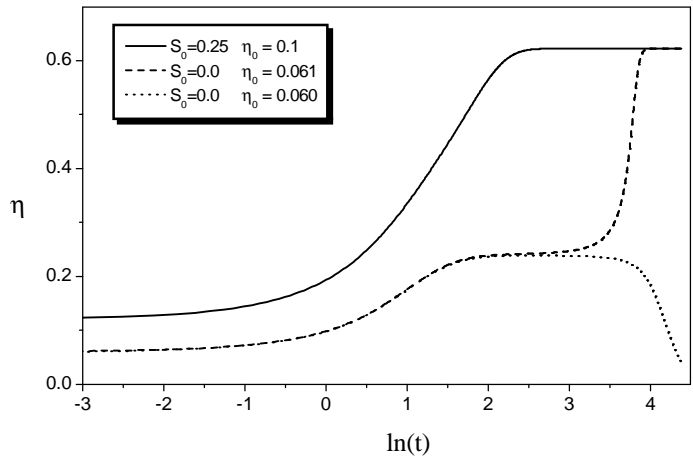
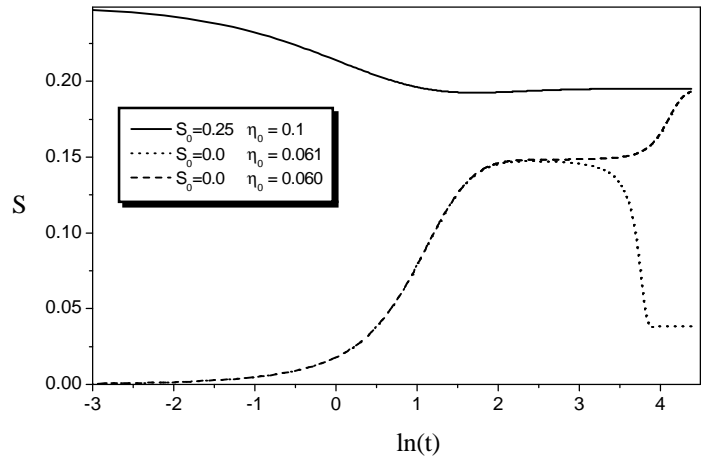
\end{document}